# Growth rate and EBSD texture analysis of nitrogen doped diamond films

T. Liu, D. Raabe *

*Max-Planck-Institut für Eisenforschung, Max-Planck-Str. 1, 40237 Düsseldorf, Germany*

* Corresponding author. Tel.: +49 211 6792 278, E-mail address: d.raabe@mpie.de (D. Raabe)

**Abstract**

Chemical-Vapor-Deposition (CVD) diamond films were prepared using a variation of nitrogen addition into the gas source admixture by a direct current CVD method. The influence of nitrogen addition on the crystallographic texture and grain shape evolution in heteroepitaxial polycrystalline diamond films was investigated using high-resolution electron backscattering diffraction (HR-EBSD) and X-ray diffraction (XRD). The analysis reveals that an addition of 1.5% $N_2$ to the $CH_4$ gas flow leads to a strong enhancement of a {110} fiber texture. The phenomenon is discussed in terms of a competitive growth selection mechanism.

Studies on diamond films are of high relevance as they are increasingly used as semiconductor in electronic devices such as ultraviolet light-emitting-diode, high-speed switches, and high-power transistors. Nitrogen-doped diamond is an N-type semiconductor while boron-doped diamond is P-type. Usually, it is more practical to synthesize N-type semiconducting Chemical-Vapor-Deposition (CVD) diamond because the high boiling point of boron prevents its incorporation into diamond. Nitrogen incorporation into the deposition atmosphere has drawbacks and benefits. Concerning the disadvantages, the existence of nitrogen reduces the thermal conductivity and the optical transparency [1-2]. In addition, nitrogen has a high solubility in diamond and is found to be either concentrated in small precipitates (type Ia) or dispersed on substitutional atomic sites (type Ib) [3]. Nitrogen addition gives diamond a yellow tarnish and increases the defect content, resulting in a negative influence on various physical properties [4]. Regarding the advantages, a small amount of nitrogen favors a <100> growth texture [3,5], which has lower roughness [6], higher wear resistance, and higher heat conductivity [7,8] than other crystallographic directions. Also, a limited incorporation of nitrogen seems to promote the deposition rate [9-11]. Thus, it is important to understand the influence of nitrogen additions in order to take advantage of the benefits and avoid the disadvantages during diamond film synthesis. Corresponding studies have been conducted using characterization by secondary-ion-mass spectroscopic [2], Raman spectroscopy [3-4], and X-ray diffraction (XRD) [3].

This paper presents an investigation on the growth rate and the crystallographic texture of diamond as a function of nitrogen additions. We used a high power direct current arc plasma jet method, operated in gas recycling mode to prepare large-scale freestanding CVD diamond films [11-13]. Sets of films were deposited using variations in nitrogen addition ranging from 0.5% to 3.5% $N_2/CH_4$. The Molybdenum (Mo) substrates were kept at 1050 ºC (±20 K) during deposition. The ratio of methane to hydrogen in the feeding flow was 1.6%. The $N_2/CH_4$ ratios in the feeding flow were set to 0.5%, 1.5%, 2.5%, and 3.5%, respectively. Since the roughness of the substrate promotes nucleation [14-15], the substrates were polished using diamond powder and cleaned in an ultrasonic bath. During the nucleation phase, a hydrogen feeding flow of 2.5% methane was applied for 15 minutes at 950 ºC (±20 K). In order to obtain excellent Kikuchi patterns, the cross-sections of all samples were cut by laser and sputtered using an $Ar^+$ ion beam to obtain a smooth surface. Microtexture analysis was conducted using HR-EBSD mapping with 1 μm step size. The growth front and the nucleation zone were also characterized using XRD pole figures.

The growth rate nearly increases by a factor of 2 (from 10.4 to 18.5 μm per hour) when the $N_2/CH_4$ ratio increases from 0.5% to 3.5%, Table 1. The highest slope is between 1.5 and 2.5% $N_2/CH_4$. The increase in growth rate as a function of the N content of heteroepitaxial diamond films is attributed to several factors. Firstly, the direct current CVD technique can achieve higher gas pressure and energy density in the plasma when N is present [13]. Secondly, N-based radicals can act as C-containing precursors during diamond synthesis [4,16]. This effect is assumed to act twofold during synthesis. The first effect is that N-based radicals can act as precursors for diamond nucleation sites. The second one is that, instead of H-based radicals, N-based radicals etch amorphous carbon phases and graphite faster than they remove diamond. It is, hence, assumed that the etching rate increases with rising nitrogen concentration to some extent, leading to an overall increase in the diamond growth rate.

Table 1. (1) Growth rate; (2) Fraction of {110} fiber texture and scatter width measured in the growth surface by XRD; (3) Fraction of {110} and {001} fiber texture and average grain diameter measured in cross sections by EBSD.

| Sample, $N_2/CH_4$ % | 0.5 | 1.5 | 2.5 | 3.5 |
|---|---|---|---|---|
| **(1)** Growth rate (μm/h) | 10.4 | 12.4 | 16.2 | 18.5 |
| **(2)** Fraction of {110} texture in growth surface, XRD | 38.5 | 79.8 | 89.4 | 73.4 |
| Scatter width (°); XRD | 32.4 | 25.9 | 35.3 | 30.5 |
| **(3)** Fraction of {110} texture in cross section, EBSD | 8.6 | 45.5 | 21.3 | 10.7 |
| Fraction of {001} texture in cross section, EBSD | 0.9 | 0.8 | 1.1 | 1.9 |
| Average grain diameter in cross section (μm) | 5 | 4.7 | 4.2 | 3.9 |

We characterized the textures in the nucleation surface (interface to substrate), the free growth surface, and the cross section for each sample. Fig. 1 shows the XRD {111} pole figure for the nucleation and growth surfaces. The nucleation textures (NS) are practically random without any significant growth competition at this early stage [17]. The textures of the free growth surfaces (GS) are more pronounced. All samples reveal {110} fiber textures, Fig. 1. The XRD pole density of the {110} texture reaches its maximum at a $N_2/CH_4$ ratio of 1.5%. The corresponding volume fractions and scatter widths of the {110} texture fibers as determined by XRD are shown in Table 1. Although the {110} volume fraction reaches its maximum at a $N_2/CH_4$ concentration of 2.5%, the scatter width for this component is lowest for 1.5% $N_2/CH_4$. Considering both, the {110} volume fraction and its scatter width, the data reveal that the orientation distribution as determined by XRD is sharpest at a $N_2/CH_4$ ratio of 1.5%.

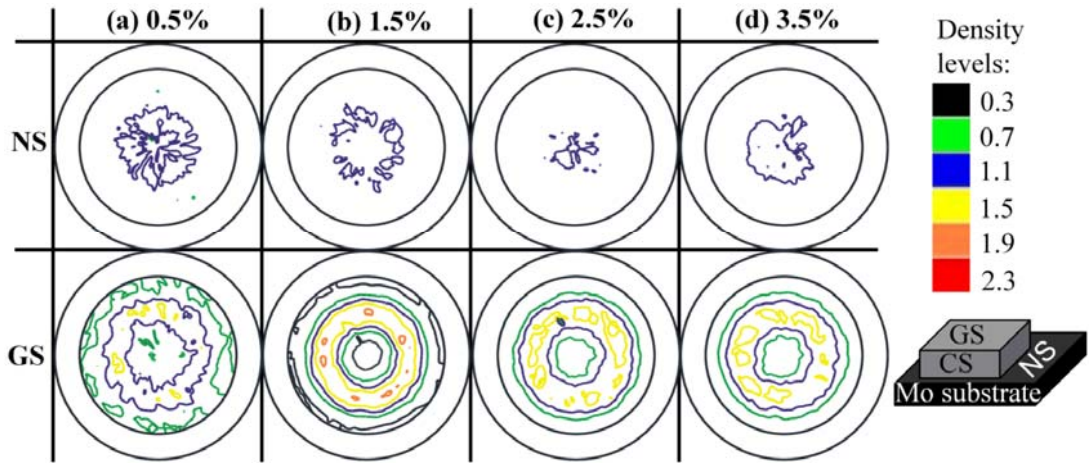

Fig. 1 {111} pole figure of the nucleation surfaces (interfaces to substrate, NS) and free growth surfaces (GS) of the four samples with different $N_2/CH_4$ ratios (CS: cross section).

Fig. 2 shows the microtexture of the cross-sections (EBSD). In each map, the bottom is the interface to the substrate and the top is the free surface. Some grains grow continuously through the film while others terminate due to growth competition and twinning [17,18]. Some of the large grains have columnar or conical shapes in growth direction. Most of the large columnar crystals belong to the {110} texture fiber (green color), Fig. 3. Small grains are typically {001} oriented (red color). Fig. 2 also shows $\varphi_2=45°$ sections of the orientation distribution functions (ODFs) obtained from EBSD analysis. These sections contain the {110} texture fiber parallel to growth direction (angles $\varphi_1$ and $\Phi$ vary between 0° and 180°) [19-22]. The data show that a $N_2/CH_4$ ratio of 1.5% leads to very pronounced and sharp {110} fiber texture (small scatter width). The maximum of the orientation density along the {110} fiber is up to 16 times random. The textures of the 3 other samples are much weaker.

The comparison of the XRD and EBSD data in table 1 deserves an explanation: Fig. 2 shows that the EBSD scans included areas of about 1000 μm × 250 μm. These scans each included a few hundred grains for each sample which represents a decent statistical ensemble. The global textures obtained by EBSD show a very similar crystallographic texture as those obtained by XRD, implying that the grain orientation distribution is fairly homogeneous in these samples. The differences in absolute numbers between the EBSD and XRD data (table 1) results in particular from the different set-up of the measurements. The XRD data were taken from the growth and substrate interfaces while the EBSD data were obtained in cross sections. The latter information is essential for obtaining detailed data on the growth competition during deposition. The XRD data cannot provide that level of lateral resolution and, hence, provide only information on the statistical texture in the surface and substrate layers.

Fig. 3 shows the aspect ratio distribution of the grains as obtained by EBSD (Fig. 2). The samples with 1.5%, 2.5%, and 3.5% $N_2/CH_4$ ratio have a high fraction of columnar and conical crystals indicating the importance of growth selection in these microstructures. The sample with 0.5% $N_2/CH_4$ ratio has a lower fraction of long conical grains which suggests that growth selection plays a minor role in this specimen. All elongated crystals with a shape ratio up to 0.25 are plotted in black in the column diagram and in the EBSD maps. The percentages below each map quantify the fraction of {110}-oriented grains within this group of conical and columnar (black) crystals. In the specimen with 1.5% $N_2/CH_4$ most of the long grains (73.9%) are {110}-oriented. For the other 3 samples the dominance of the {110} texture among the conical and columnar grains is much less pronounced. This observation suggests that the strong {110} fiber texture in the sample with 1.5%N$_2$/CH$_4$ is connected to a growth selection mechanism. This argumentation is based on the topological fact that elongated conical grain shapes can develop during competitive thin film growth only by overgrowing neighbor crystals that grow slower. In the case of the 1.5%N$_2$/CH$_4$ sample these fast growing crystals are not randomly oriented but they are essentially {110} oriented (73.9%) so that we conclude a connection between growth selection and nitrogen content.

Also, the EBSD data show that the average grain diameter of the four samples decreases from 5 to 3.9 μm when the N$_2$/CH$_4$ ratio rises from 0.5% to 3.5%, table 1. The average grain diameter is here expressed as the equivalent diameter of a circle with the same area. Although this characterization does not consider the grain shape, it gives a relative measure for the size of the diamond grains.

The EBSB analysis also revealed a high {110} oriented volume fraction (scatter width 10°) of 45.5% for a N$_2$/CH$_4$ ratio of 1.5% but much smaller values for the other samples as shown in Table 1 and Fig. 2. Since the averaged grain size (expressed in terms of the equivalent diameter) does not show an analogous tendency we conclude that the formation of the {110} texture cannot be attributed to regular grain growth during deposition but to growth selection as also suggested by the conical elongated grain shapes observed in Figs. 2 and 3. A more detailed analysis of the texture data reveals that a rising nitrogen concentration promotes the {001} fiber texture, table 1. The rising nitrogen concentration, thus, seems to slightly promote a {001} texture which does not fully conform with earlier work [3,5].

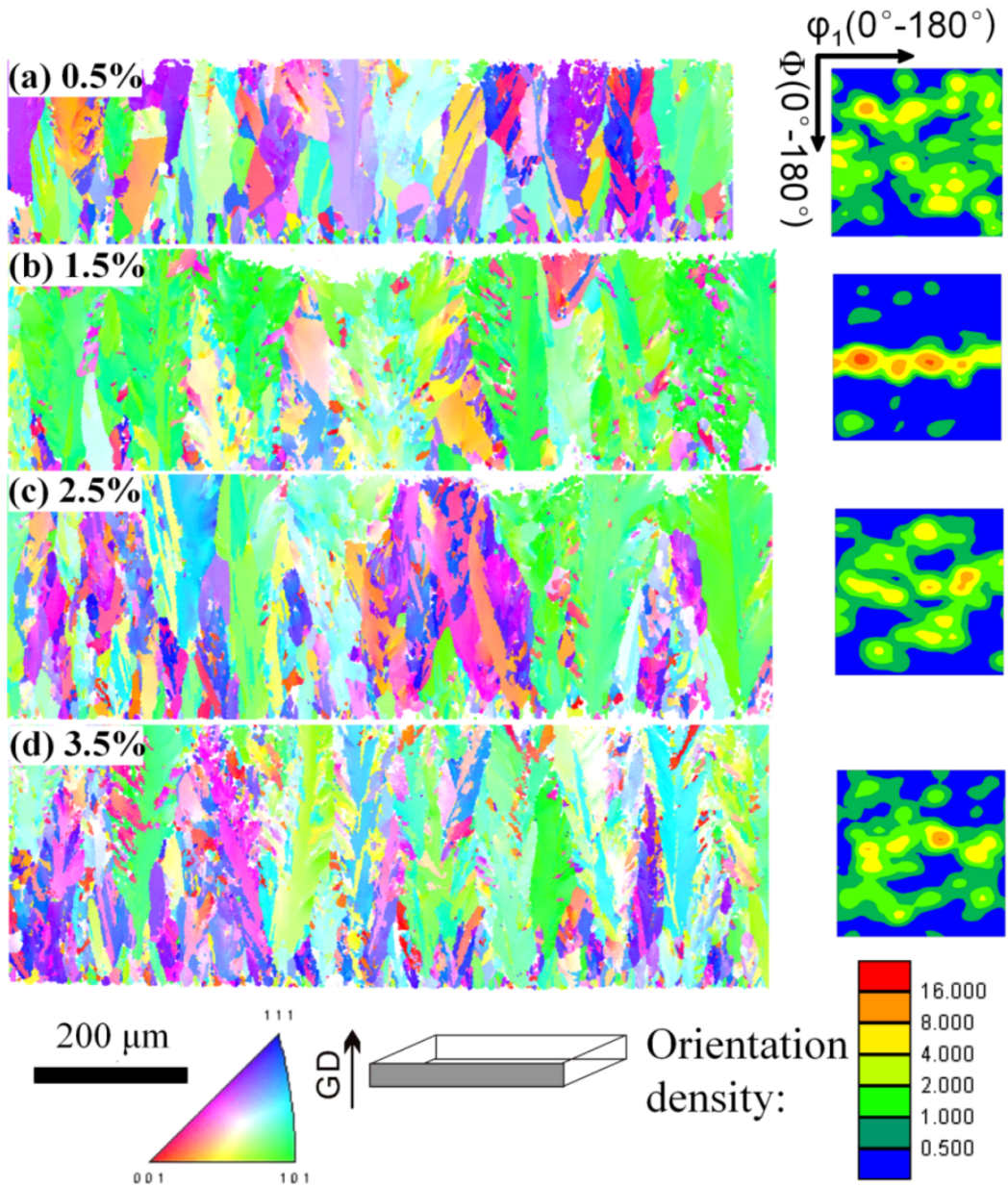

Fig. 2 Orientation maps and EBSD-based ODF sections at $\varphi_2 =45°$ of the four samples. 0.5%, 1.5%, 2.5%, and 3.5% $N_2/CH_4$. Orientation maps are color coded for the axis parallel to the growth direction.

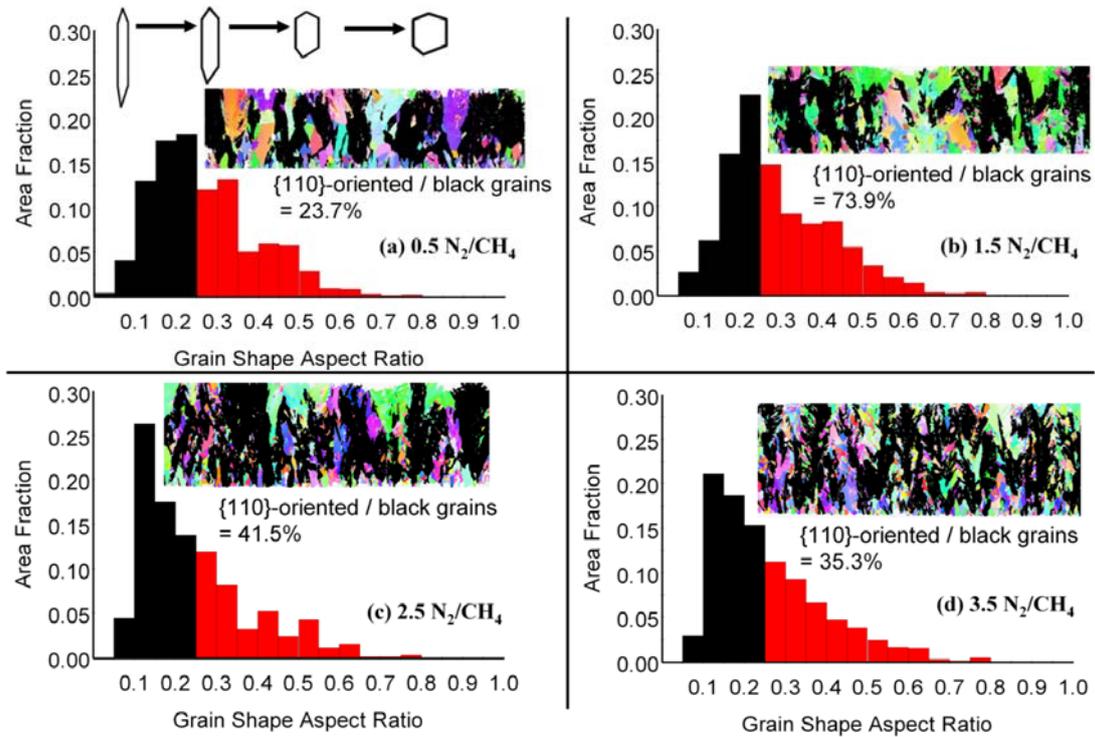

Fig. 3 Distribution of the grain shape aspect ratios in the four diamond samples. The black columns below 0.25 correspond to all grains that are highlighted in black in the inverse pole figure maps (compare to Fig. 2). The fraction of the {110}-oriented grains relative to all black grains (characterized by a long aspect ratio) reach from 23.7% to 73.9% (orientation scatter 15°). Grains on the edge of the map were considered as regular crystals and included in the analysis.

In summary, the film growth rate increases from 10.4 μm to 18.5 μm per hour with rising $N_2/CH_4$ ratio (from 0.5% to 3.5%) owing to the effect of the rising nitrogen concentration on the diamond growth kinetics. The average equivalent grain diameter of the four film samples decreases from 5 μm to 3.9 μm with increasing nitrogen content. {110}-oriented grains are often associated with an aspect ratio below 0.25 in all four samples. For 1.5% $N_2/CH_4$ even 73.9% of all columnar crystals have a {110} orientation indicating that a growth selection mechanism prevails. In this specimen the {110} volume fraction reaches 45.5% of the overall texture.

The authors acknowledge the support of the International Max Planck Research School for Surface and Interface Engineering in Advanced Materials (IMPRS-SurMat), and the supply of the diamond films from Mr. Hui Guo of Hebei Academy of Sciences, China.